\documentstyle[11pt,newpasp,twoside,epsf]{article}
\def\simlt{\lower.5ex\hbox{$\; \buildrel < \over \sim \;$}}
\def\simgt{\lower.5ex\hbox{$\; \buildrel > \over \sim \;$}}
\def\kms{km s$^{-1}$}
\def\vlsr{v_{\rm LSR}}

\def\schi{{\sc Hi}\ }

\def\nhi{{N({\rm HI})}}
\def\nhii{{N({\rm H_2})}}

\def\edcomment#1{\iffalse\marginpar{\raggedright\sl#1\/}\else\relax\fi}
\reversemarginpar

\index{supernova remnants}
\index{supernova remnants!HI 21cm line}
\index{supernova remnants!G347.3--0.5}
\index{supernova remnants!surveys}
\index{supernova remnants!interactions with molecular clouds}

\begin{document}
\title{{\sc Hi} Study of Southern Galactic Supernova Remnants}
\author{Bon-Chul Koo and Ji-hyun Kang}
\affil{SEES, Seoul National University, Seoul 151-742, Korea}
\author{Naomi McClure-Griffiths}
\affil{ATNF-CSIRO, PO Box 76, Epping NSW 1710, Australia}

\begin{abstract}
We briefly summarize the survey of \schi 21 cm emission lines 
to search for shocked atomic gas associated
with Galactic supernova remnants (SNRs) in the southern sky.
For G347.3$-$0.5, we discuss the distance to the SNR and the implications 
of the \schi results.

\end{abstract}

\section{The Survey}

We carried out a \schi 21-cm line survey to search for shocked atomic gas associated 
with Galactic SNRs in the southern sky. 
We have studied 97 SNRs between $\ell=253^\circ$--$358^\circ$  in the Green's catalog 
using the Southern Galactic Plane Survey (SGPS) data.
We compare the average \schi spectra of SNRs to those of the surrounding regions, and
look for excessive emission wider than 10~\kms\ and
localized at the position of the SNRs.
We divide the SNRs into 3 ranks,
where the increasing number implies 
increasing reliability of a detected \schi feature. 
The ranks correspond to those of Koo \& Heiles (1991), who did
a similar survey toward the northern SNRs. 
Of the 97 SNRs, 10 SNRs are ranked as 3, 22 SNRs are ranked as 2,
and others are ranked as 1. 
Rank 3 SNRs have very high-velocity \schi gas 
confined to the SNR, and the association is quite plausible.

\section {G347.3$-$0.5}

G347.3$-$0.5 (RX J1713.7$-$3946) is one of the rank 3 SNRs. 
It is a shell-type SNR with 
X-rays dominated by non-thermal synchrotron radiation like SN 1006.
TeV gamma-ray emission has been detected toward the X-ray bright
western shell (Muraishi et al. 2000).
There are two different interpretations for G347.3$-$0.5:
(1) it is a young SNR at a distance of 1 kpc, possibly the remnant of the SN AD 393 
(Koyama et al. 1997), and (2) it is an old SNR at a distance of 6 kpc, 
possibly interacting with molecular clouds (Slane et al. 1999).

\begin{figure}[t]
\hfill
\epsfxsize=0.52\textwidth \epsfbox{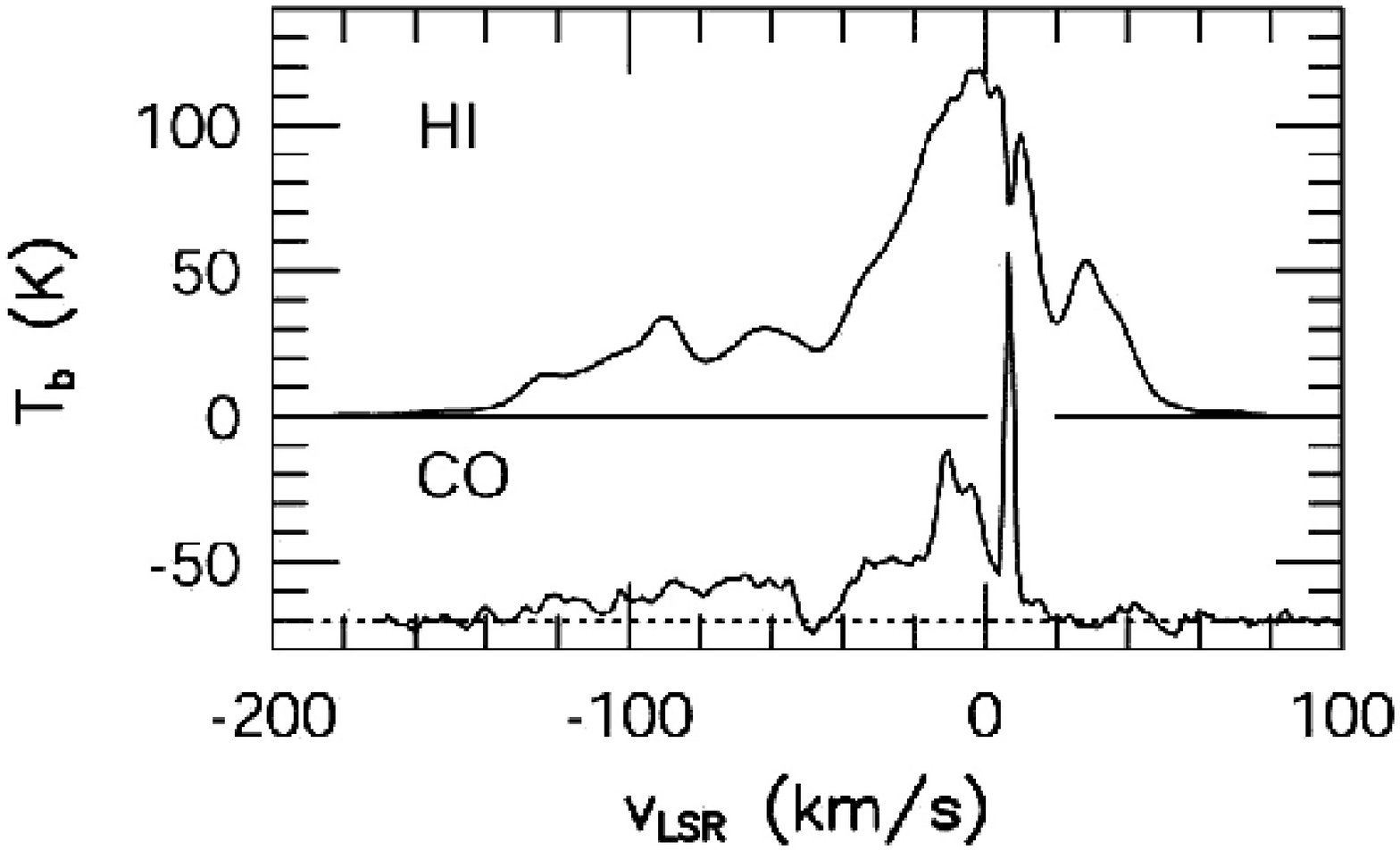}
\epsfxsize=0.50\textwidth \epsfbox{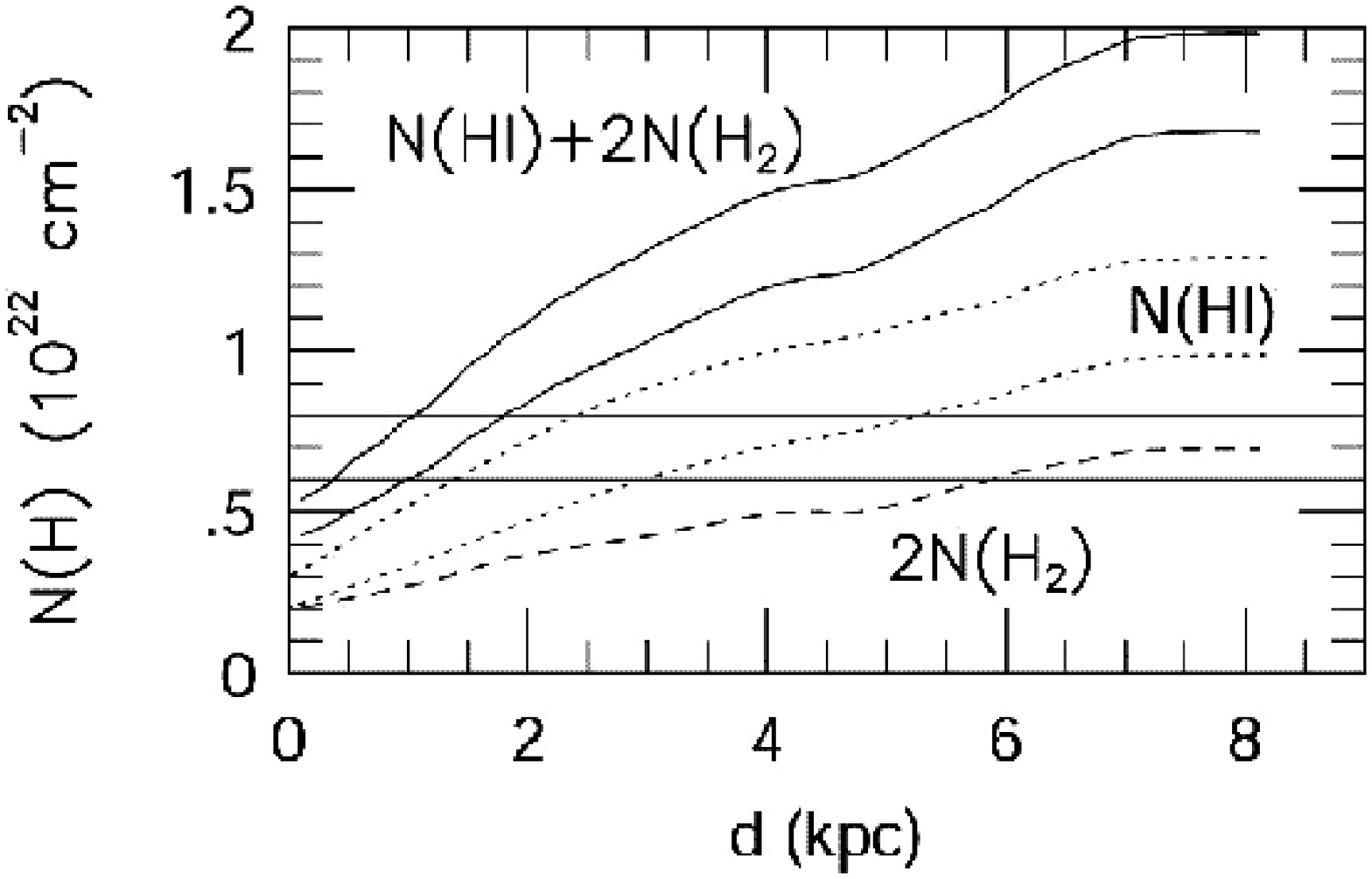}
\hfill
\caption{(left) \schi and CO (J=1--0) 
profiles toward the SNR. The peak intensity of the CO emission is 1.3 K.
(right) Hydrogen column density as a function of distance toward the direction
of G347.3$-$0.5. For $\nhi$ (and $\nhi+2\nhii$), the upper and lower graphs correspond to 
spin temperatures of 130~K and 200 K, respectively.
}

\end{figure}

Fig. 1 (left) shows average \schi and CO profiles toward the direction of the SNR.
The \schi profile is obtained from the SGPS data while the CO profile is 
from the CfA CO survey (Dame et al. 2001).
The total \schi column density along the line of sight 
is (1.8--2.3)$\times 10^{22}$~cm$^{-2}$ assuming a constant 
spin temperature of 130--200~K.
The total column density of H$_2$ molecules, using  
$\nhii/I({\rm CO})=1.9\times 10^{20}$cm$^{-2}$ (K \kms)$^{-1}$, is
$\simeq 0.4 \times 10^{22}$~cm$^{-2}$. Hence,
the total column density of H nuclei 
is $(2.6-3.1)\times 10^{22}$~cm$^{-2}$,
which is considerably greater than 
the X-ray absorbing columns  
($\simeq 0.6$--$0.8\times 10^{22}$~cm$^{-2}$ Slane et al. 1999).
Fig. 1 (right) shows how 
the column densities vary as a function of distance, which was obtained by 
converting the LSR velocity to distance assuming a flat Galactic rotation
model. There is a distance ambiguity, and we arbitrarily assumed that 
2/3 (1/3) of the column density at a given velocity 
is from the gas on the near (far) side. 
We also assumed that 2/3 of the \schi emission between 
$\vlsr=0$ and +10~\kms\ is from the local gas, 
while all the H$_2$ emission in this velocity range
is assumed to be from the local gas because there is a self-absorption 
\schi feature associated with the molecular gas.  
According to Fig. 1 (right), the distance at which the accumulative 
column density equals the X-ray absorbing columns is $\sim 1$ kpc.

The excess \schi emission toward G347.3$-$0.5 is detected between
$\vlsr=68$ and $81$~\kms\ at the western rim where the X-rays are enhanced.
At a distance of 1 kpc, the radius ($30'$) of the SNR is 9 pc and 
the detection of shock-accelerated \schi gas 
implies an interaction with a dense medium.
Previous studies showed that the fast-moving \schi gas is often produced by 
SNR shocks propagating through molecular clouds, and 
G3473.$-$0.5 may interact with molecular clouds too.

\acknowledgements
This work was supported by S.N.U. Research Fund (3345-20022017).

\begin {references}
\reference Dame, T. M., Hartmann, D., \& Thaddeus, P. 2001, ApJ, 547, 792
\reference Koo, B.-C., \& Heiles, C. 1991, ApJ, 382, 204
\reference Koyama, K. et al. 1997, PASJ, 49, L7
\reference Muraishi et al. 2000, A\&A, 354, L57
\reference Slane, P. et al. 1999, ApJ, 525, 357
\end{references}

\end{document}